\documentclass[a4paper,10pt]{aa}
\usepackage{amsmath}
\usepackage{epsfig}
\usepackage{graphicx}

\begin{document}

\authorrunning{E. van der Swaluw et al.}
\title{Convective magneto-rotational instabilities in accretion disks}
\author{E. van der Swaluw \inst{1}, J. W. S. Blokland \inst{1,2}, \and R. Keppens \inst{1,2}}
\institute{
FOM-Institute for Plasma Physics Rijnhuizen, P.O. Box 1207, 3430 BE Nieuwegein, The Netherlands
\and
Association EURATOM-FOM, Trilateral Euregio Cluster
} 
\offprints{E. van der Swaluw,
\email{swaluw@rijnh.nl}}
\date{}

\abstract{
We present a study of instabilities occuring in thick magnetized accretion
disks. We calculate the growth rates of these instabilities and characterise
precisely the contribution of the magneto-rotational and the convective mechanism.
All our calculations 
are performed in radially stratified disks in the cylindrical limit. 
The numerical calculations are performed using the appropriate local 
dispersion equation solver discussed in Blokland et al. (\cite{BSKG}).
A comparison with recent results by Narayan et al. (\cite{NQIA}) shows
excellent agreement with their approximate growth rates only if the disks are weakly 
magnetized. However, for disks close to equipartition, the dispersion
equation from Narayan et al. (\cite{NQIA}) loses its validity.
Our calculations allow for a quantitative determination of the increase of the growth rate
due to the magneto-rotational mechanism. We find that the increase of the growth 
rate for long wavelength convective modes caused by this mechanism is almost neglible. 
On the other hand, the growth rate of short wavelength instabilities can be  
significantly increased by this mechanism, reaching values up to 60\%.
}
\titlerunning{Convective magnetorotational instabilities in magnetized accretion disks}
\maketitle
 
\section{Introduction}

Accretion disks are present 
around a variety of astrophysical objects, ranging from young proto-stars 
in star formation regions to massive black holes in the centers of 
galaxies. 
A standard model of a geometrically thin  accretion disk was introduced 
in the early seventies by Shakura \& Sunyaev (\cite{SS}). In their model angular 
momentum is transported outwards by an anomalous viscosity mechanism, which is 
parametrised by the $\alpha$-parameter. This parameter scales linearly with the 
radial-azimuthal component of the stress tensor, which is associated with the 
viscous torque providing the angular momentum transport. The assumption of a 
geometrically thin disk implies that the gravitational energy release
by viscous dissipation is locally radiated away.

It was realised in the early nineties that the anomalous viscosity mechanism
can be provided by the turbulence arising from the magneto-rotational instability
(Balbus \& Hawley \cite{BH91}). These authors showed that this essentially magnetic
instability is
a very robust one, occuring in weakly magnetized thin accretion disks. 

However, in recent years, Chandra observations have found examples of
underluminous black holes at X-ray frequencies. A good example is our 
own Galactic Center, Sgr A$^{*}$ (Baganoff et al. \cite{Bag01}), and
the galactic center of the elliptic galaxy M87 (Di Matteo et al. \cite{DiMat}). 
These observations
might be explained in the context of nonradiating accretion flows, which are
still using the $\alpha$ description from Shakura \& Sunyaev (\cite{SS}), but 
are no longer geometrically thin (Narayan \& Li \cite{NY}). One of these models 
which has gained a lot of interest in the literature in recent years is the 
convection-dominated accretion flow (CDAF). In these type of accretion flow models, 
the transport of the angular momentum outwards by the turbulence arising from the
magneto-rotational instability is partly counterbalanced by transport inwards due to 
convective motion occuring in these {\it thick} accretion disks. Therefore the 
accretion rate might be efficiently reduced, which might explain the lower observed
X-ray luminosities in Sgr A$^*$ and the central region of M87.

Recent studies of these CDAF-type of disks, in which both convective and magneto-rotational
instabilities can be found, have been performed by Balbus \& Hawley (\cite{BH02})
and Narayan et al. (\cite{NQIA}). These authors have used a linear stability 
analysis and identified unstable long-wavelength modes with a convective nature,
and short-wavelength modes with a more dominant MRI behaviour.

In this paper, we will elaborate on the work from Narayan et al. (\cite{NQIA}).
These authors have considered a differentially rotating and thermally stratified 
plasma, with a weak axial magnetic field. They use the equation for the growth 
rate as obtained by Balbus \& Hawley (\cite{BH91}), who used linear stability theory 
to derive the latter. Narayan et al. (\cite{NQIA}) consider five models of an accretion 
disk, each of these five models has a different value of the assumed uniform 
Brunt-V\"ais\"al\"a frequency.
Their results clearly show the rise of a plateau in the obtained growth rate at low 
axial wavenumbers as the 
Brunt-V\"ais\"al\"a frequency is increased. They argue that the modes in this 
plateau are identified as convective modes, once the Brunt-V\"ais\"al\"a frequency 
exceeds the epicyclic frequency, i.e. the H\o iland criterion (Tassoul, \cite{T87}). 
The exact nature of a mode in this plateau is identified using results from the 
linear stability theory as performed by Balbus \& Hawley (\cite{BH91}, \cite{BH02}).  

We follow up on the above mentioned work, but we will numerically solve a sixth 
order polynomial dispersion relation, using the Local Dispersion Equation Solver 
(LODES), which was already discussed in Blokland et al. (\cite{BSKG}). We will 
not only consider weakly magnetized disks, but also consider models which are 
closer to equipartition. We will consider a differentially rotating plasma for which the equilibrium quantities
are power-law profiles of the radius. Furthermore, we explicitly define a value for 
the scale height of the disk $H$ with respect to the radius $r$, with the free parameter
$\epsilon=H/r$. 

In our present analysis we only consider axisymmetric perturbations, which means that
in a model with a purely toroidal magnetic field, the magneto-rotational mechanism
is excluded. Therefore in such models, the obtained modes are purely convective.
The growth rates from these modes can be compared to a similar model which
includes a {\it weak} axial magnetic field. This enables one to observe the 
increase of the growth rate due to the presence of the magneto-rotational 
mechanism. 

This paper is organised as follows: in section 2 we discuss our model, section 3 
shortly recalls the model from Narayan et al. (\cite{NQIA}), in section 4 we
present our results, including a comparison with the results from Narayan et al. 
(\cite{NQIA}), finally in section 5 we present our conclusions.

\section{Accretion disk model}

\subsection{Equilibrium of a differentially rotating plasma}

We want to investigate the growth rate of instabilities occuring in 
a magnetized accretion disk. In order to quantify instabilities using a 
linear analysis, one has to consider an MHD equilibrium. In our case, this
equilibrium is one of a differentially rotating plasma, which is radially 
stratified. In our model the density, pressure, magnetic field strength and 
toroidal velocity only depend on the radius $r$. This type of model is sometimes 
referred to as an accretion disk in the cylindrical limit (see for example
Hawley \cite{Hawley01}). 

We use an equilibrium as in Blokland et al. (\cite{BSKG}), generalised with an
additional parameter $a$, which is introduced in order to allow for convective
instabilities. The following profiles are used for respectively
density $\rho$, thermal pressure $p$, toroidal magnetic field $B_{\theta}$,
axial magnetic field $B_{\rm z}$ and the toroidal velocity $v_{\theta}$:
\begin{eqnarray}
\rho & = & r^{-(3+a)/2}, \\
p  & = & \epsilon^2 \;\; r^{-(5+a)/2}, \\
B_{\theta} & = & -\alpha_1
\sqrt{{2\epsilon^2\over\beta(\alpha_1^2+\alpha_2^2)}}\;\; r^{-(5+a)/4}, \\
B_z & = & \alpha_2
\sqrt{{2\epsilon^2\over\beta(\alpha_1^2+\alpha_2^2)}}\;\; r^{-(5+a)/4}, \\
v_\theta  & = &  V_0\;\; r^{-1/2}, \\
\end{eqnarray}
in which the $\alpha$-parameters can be expressed as the ratio of 
both the toroidal and axial magnetic field over the total magnetic field
$B$ like:
\begin{eqnarray}
B_\theta/B\; &=& -\alpha_1/\sqrt{\alpha_1^2 + \alpha_2^2},\; \\
B_z/B\; &=& \alpha_2/\sqrt{\alpha_1^2 + \alpha_2^2}\; ,
\end{eqnarray}
and the parameter $V_0$ is defined as:
\begin{equation}
V_0^2 =  GM_\ast - {\epsilon^2\over 2\beta(\alpha_1^2 + \alpha_2^2)}
\left((5+a)(1+\beta)(\alpha_1^2 + \alpha_2^2) - 4\alpha_1^2\right).
\end{equation}
Here, $G$ and $M_\ast$ denote respectively  the gravitational constant 
and the mass of the central object. Finally, the parameter $\beta$ 
denotes the radially constant ratio between the thermal pressure and the total magnetic 
pressure $B^2/2$, i.e. $\beta\equiv 2p/B^2$.

The above power-law profiles satisfy the radial force balance equation 
in cylindrical coordinates:
\begin{equation}
  \label{eq:forcebalance}
  \left[ p + \tfrac{1}{2}B^{2} \right]' + \frac{B_{\theta}^{2}}{r} = \frac{\rho
v_{\theta}^{2}}{r} - \rho g\; ,
\end{equation}
here $g$ denotes $GM_\ast /r^2$, and the prime indicates the derivative with
respect to the radius $r$.

The differences between our equilibrium and the one from Narayan et al. 
(2002) are: 1) we {\it explicitly} define the above mentioned equilibrium quantities by using 
power-law profiles as a function of radius $r$; 2) we include the toroidal magnetic 
field component $B_\theta$; 3) we define a ratio of the scale height $H$ over the 
radius $r$, i.e. $\epsilon\equiv H/r$, in order to quantify the thickness of our disk model; 
and 4) we do not exclude disks close to equipartition ($\beta$ is a free parameter).

\subsection{Determining the growth rate of instabilities}

We use the local dispersion equation solver (LODES), which was 
recently discussed by Blokland et al. (\cite{BSKG}). This code
calculates the growth rate of the most unstable mode in a given MHD
equilibrium for a given radial `wavenumber' ($q$), and a toroidal 
($m$) and axial wavenumber ($k$) at a given position $r_{\rm i}$. 
We will only consider axisymmetric perturbations ($m=0$) in this 
paper. 

In order to determine the radial `wavenumber' $q$, we use the  
method as discussed by Blokland et al. (\cite {BSKG}). It was shown
there that under certain assumptions, the numerical solution of the
full set of linearised compressible MHD equations governing all MHD modes in 
disk equilibria obeying Eq. (10), can be avoided for modes obeying a local
dispersion equation found from WKB analysis. Excellent agreement between
the growth rate and the eigenfunction behaviour was demonstrated when
position $r_{\rm i}$ and associated `wavenumber' $q$ were properly
calculated from the full numerical solution. In this paper we follow
Balbus \& Hawley (\cite{BH91}) and Narayan et al. (\cite{NQIA}), who
effectively take $q=0$, in order to make a direct comparison with their
calculations. We notice that we only observed a difference of the order 
of 1\% in our obtained growth rates, with respect to our calculations for 
which the `wavenumber' $q$ was properly calculated.
 

The local dispersion equation solver finds the root of a sixth order 
polynomial, using Laguerre's method (Press et al. \cite{PTVF}).
This sixth order polynomial dispersion equation represents the local
dispersion equation as an approximation to the true $10^{\rm th}$ degree
WKB local dispersion equation (Blokland et al., \cite{BSKG}), which is 
more general than the one discussed by 
Narayan et al. (\cite{NQIA}):
1) it allows to calculate the growth rate of instabilities for an equilibrium
which has both an axial and a toroidal magnetic field component; and
2) the equilibrium can be either weakly magnetized, or close to equipartition.
Narayan et al. (\cite{NQIA}) could not consider disks close to equipartition,
because the fourth order polynomial from Balbus \& Hawley (\cite{BH91}) 
was derived {\it assuming} a weak axial and no toroidal magnetic field 
component.

\begin{figure}
\resizebox{\hsize}{!}{\includegraphics{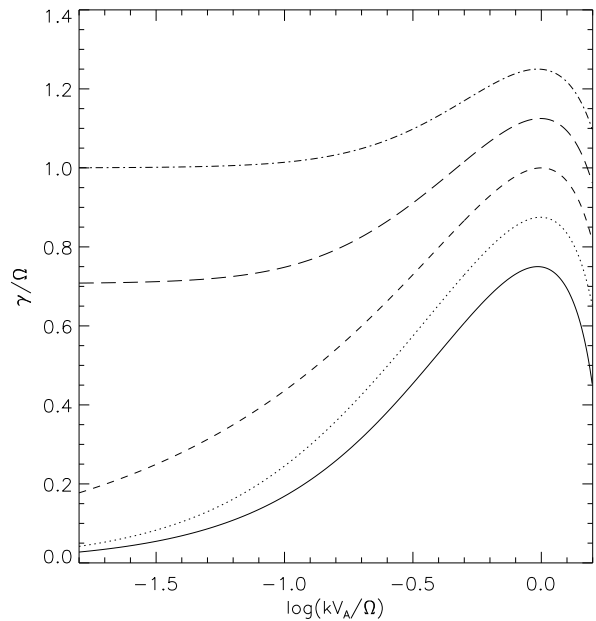}}
\caption{Dimensionless growth rate $\gamma/\Omega$ as a function of 
dimensionless wavevector $kv_{\rm A}/\Omega$ for accretion disk models
from Narayan et al. (\cite{NQIA}). The five growth rates shown are from
five different solutions: $N^2/\Omega^2=0.0$ (solid line); 
$N^2/\Omega^2=0.5$ (dotted line); $N^2/\Omega^2=1.0$ (short-dashed line);
$N^2/\Omega^2=1.5$ (long-dashed line); $N^2/\Omega^2=2.0$ (dot-dashed line).} 
\label{Pap2Fig1}
\end{figure}

\section{Accretion disks with convection}

\subsection{A thick accretion disk}

One of the key parameters of our MHD equilibrium is the parameter $\epsilon$,
which is taken as a constant free parameter. This parameter was introduced by
Shakura \& Sunyaev (\cite{SS}) in their model of geometrically thin accretion 
disks, which are Keplerian rotating. The physical interpretation of $\epsilon$
in these models is the same as in the MHD equilibrium we consider here, i.e. 
$\epsilon=c_{\rm s}/v_{\theta}\simeq H/r$, were $c_{\rm s}$ denotes the sound
speed. MRI instabilities in thin accretion 
disk models were considered by Blokland et al. (\cite{BSKG}). However, in this 
paper we consider values of $\epsilon\sim 1$, i.e. we are in the regime of 
sub-Keplerian rotating disks (see e.g. Narayan \& Yi \cite{NY}). For these cases 
we stick to the interpretation $\epsilon\simeq H/r$, noting that the identification 
$\epsilon=c_{\rm s}/v_{\theta}$ is not valid anymore. Typically, $c_{\rm s}/v_{\theta}$ 
in our models will be larger up to a maximum factor of order $\sim 10$.
As mentioned in section 2, our MHD equilibrium is an example of an accretion disk model
in the cylindrical limit. One has to realise that results from these type of models
approximate the {\it interior} of a real height-dependent accretion disk. This 
approximation is correct as long as the axial wavenumber $k$ is much larger than the 
inverse of the scale height $H$. Therefore the identity $k\gg 2\pi /(\epsilon r)$ has to 
be satisfied in our calculations in order to connect these results with the interior of 
an accretion disk.

\subsection{Convective instabilities}

As discussed in section 2, Narayan et al. (\cite{NQIA}) also considered an 
equilibrium of a differentially rotating plasma, which only allows for a 
weak axial magnetic field component. Furthermore, their equilibrium quantities 
depend on radius $r$, but are not explicitly defined as power-laws obeying
the radial force balance equation (10). 
Therefore, in order to introduce convective instabilities in their model,
they describe the stratification of the differentially rotating plasma in 
terms of a free parameter $N^2$, which is directly related to the
Brunt-V\"ais\"al\"a frequency $N_{\rm BV}$:
\begin{equation}
N^2\; =\; -\; N_{\rm BV}^2\; =\; {3\over 5\rho}\; {{\rm d} p\over {\rm d} r}\;
{{\rm d}\; {\rm ln}(p\rho^{-5/3})\over{\rm d} r}\; .
\end{equation}
In a weakly-magnetized rotating medium, the onset of convection is 
determined by the H\o iland criterium, i.e. 
$N^2 > \kappa^2(\equiv 2v_\theta(rv_\theta)'/r^2)$, where 
$\kappa$ is the epicyclic frequency. Furthermore, they assume a purely
Keplerian system, which means that $\kappa^2 = \Omega_{\rm K}^2$, therefore convective
instabilities will be present in their model when $N^2 > \Omega^2$.
Narayan et al. (\cite{NQIA}) present growth rates of instabilities 
as a function of the axial wavenumber. They use the equation for the 
growth rate as derived by Balbus \& Hawley (\cite{BH91}). They normalise their
obtained growth rates to the rotational frequency, furthermore the wavevector is 
multiplied by the Alfv\'en velocity $V_{\rm A}$, and also normalised to the rotational 
frequency. In this way the only free parameter left for the growth rate profiles is 
the parameter $N^2/\Omega^2$. Figure 1 is showing these scaled growth rates for the different 
values of $N^2/\Omega^2$ as considered by Narayan et al. (\cite{NQIA}).

In our cylindrical accretion disk model with power-law equilibrium
profiles one can calculate $N^2/\Omega^2$,
and make a direct comparison with the work of Narayan et al. 
(\cite{NQIA}). Furthermore, it can be shown that in our MHD equilibrium the
equality $\kappa^2=\Omega^2$ is always valid, even if the system is not Keplerian 
rotating (i.e. when $\epsilon\sim 1$).  Therefore our MHD equilibrium will also be convectively
unstable once $N^2 > \Omega^2$, like in the model from Narayan et al. (\cite{NQIA}), 
the difference being that our models captures significant deviations from 
{\it Keplerian rotating} disks by means of the $\epsilon$ parameter $(\Omega^2\ne\Omega_{\rm K}^2)$.

The H\o iland criterium is valid for a hydrodynamical equilibrium, but strictly 
speaking it is not valid for an MHD 
equilibrium which is weakly magnetized. However, it can be used as an indication for
convective instabilities at those wavenumbers where the restoring magnetic tension force 
is much smaller than the buoyancy force (see also Christodoulou et al., \cite{CCK}).
These last authors derived a stability criterium for axial perturbations for magnetized
equilibria with purely toroidal magnetic fields. This criterium is valid for all values
of the plasma parameter $\beta$, which therefore includes equilibria which are weakly
magnetized or close to equipartition. Therefore this criterium could be used to determine 
the presence of convective instabilities in our models with purely toroidal magnetic fields.
The criterium can also be derived from the local dispersion equation 
(see e.g. Keppens et al. (\cite{KCG02}); Blokland et al. (\cite{BSKG}); Wang et al. (\cite{WBKG})).

\begin{figure}
\resizebox{\hsize}{!}{\includegraphics{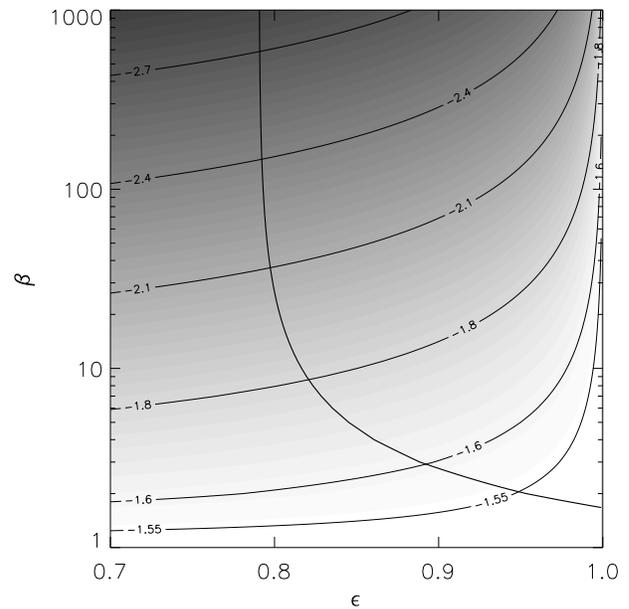}}
\caption{Logarithmic gray-scale plot of the value ${\rm log}(k_{\rm min}
V_{\rm A}/\Omega)$ in the model for which $\alpha_1=300.0, \alpha_2=1.0$
and $a=-3.0$. The solid thick line corresponds with $N^2=\Omega^2$. The
thick solid curve for the H\o iland criterium deviates from a straight vertical 
line, because of the {\it dynamical importance} of the magnetic field on the 
equilibrium for low values of $\beta$.} 
\label{Pap2Fig2}
\end{figure}

\subsection{A model for the interior of a thick accretion disk}

In our MHD equilibrium, the parameter $\epsilon$ is a free parameter
to be interpreted as related to the scale-height $H$ at each
radial position $r$. In order to perform a meaningful stability calculation
for an accretion disk, the wavenumbers of the unstable modes considered
must be such that they are able to manifest
themselves in the {\it interior} of the accretion disk. These instabilities
{\it fit} into the interior of the considered accretion disk model once the
corresponding axial wavenumber
$k > k_{\rm min}\equiv{2\pi /H}$. The analysis from Narayan et al. (\cite{NQIA}) 
shows a maximum of the growth rate at ${\rm log} (kV_{\rm A}/\Omega)\sim 0.0$
(see Figure~\ref{Pap2Fig1}). We therefore choose our MHD equilibria such that
the numerical value of ${\rm log} (k_{\rm min}V_{\rm A}/\Omega)$ is typically 
less then $\sim -1.5$. For these models, all unstable modes found with an axial 
wavenumber $k > k_{\rm min}$ {\it fit} into the interior of the considered disk 
model, and include the most unstable ones.

We found that in order to satisfy the restriction $k > k_{\rm min}$ with 
${\rm log} (k_{\rm min}V_{\rm A}/\Omega)\sim -1.5$ a high ratio of the toroidal magnetic 
field strength to the axial magnetic field strength is needed. 
Figure~\ref{Pap2Fig2} shows the parameter ${\rm log}\; (k_{\rm min}V_{\rm A}/\Omega)$ 
as a function of the parameters $\epsilon$ and $\beta$ of our model, for which the other parameters 
have been fixed. Indeed for most values of $\epsilon$ and $\beta$ the corresponding numerical value 
${\rm log}\; (k_{\rm min}V_{\rm A}/\Omega) < -1.5$. 
The thick solid line in Figure~\ref{Pap2Fig2} marks the H\o iland criterium, i.e. $N^2=\Omega^2$. The 
models on the right hand side of this line are convectively unstable, the models on the left hand side 
are convectively stable. Notice that the line $N^2=\Omega^2$ can be drawn uniquely as a function
of $\beta$ and $\epsilon$ once $\alpha_1$ and $\alpha_2$ are determined. This is because the numerical
value of $N^2/\Omega^2$, calculated from the equilibrium presented in section 2, does not depend
on the radius $r$.

\begin{figure}
\resizebox{\hsize}{!}{\includegraphics{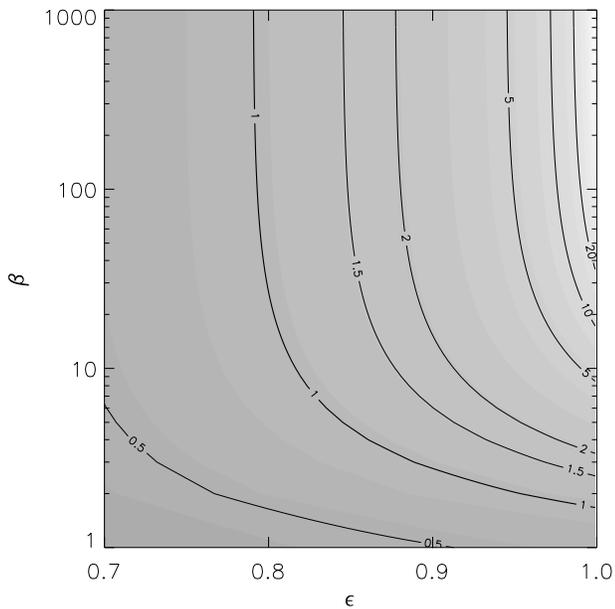}}
\caption{Gray-scale plot of the parameter $N^2/\Omega^2$ in the model
for which $\alpha_1=300.0, \alpha_2=1.0$ and $a=-3.0$.} 
\label{Pap2Fig3}
\end{figure}

As argued above we will choose a configuration of an accretion disk with most of the
magnetic field strength in the toroidal component, in order to allow for both convective and
magneto-rotational instabilities. The values of the ratio of the Brunt-V\"ais\"al\"a 
frequency and the rotational frequency marks the free parameter in the paper from 
Narayan et al.(\cite{NQIA}). We plot the numerical value of this parameter in our models 
in figure~\ref{Pap2Fig3}. We have plotted contours with $N^2/\Omega^2=0.5,1.0,1.5,2.0$, 
corresponding to the models discussed in Narayan et al.(\cite{NQIA}). It can be noted that in 
very weakly magnetized, thick disk models 
we obtain high values for $N^2/\Omega^2$, the highest value being 
$N^2/\Omega^2=600$ (model A in Table 1). Those are expected to be
rigorously convective in a pure hydrodynamical sense. 
 
\section{Accretion disks with two instabilities}

\subsection{Introduction}

We present the results from a linear analysis performed for different MHD equilibria,
using the local dispersion equation solver (LODES). In our calculations 
we will only consider equilibria with a constant density ($a=-3.0$), while most of the magnetic field 
strength is in the toroidal direction ($\alpha_1=300.0;\;\alpha_2=1.0$). Finally, we will 
only consider thick accretion disk models by considering MHD equilibria with  
$\epsilon \ge 0.8$, in order to allow for mixed convective magneto-rotational instabilities. 
We will consider five different MHD equilibria and calculate the growth rates of the unstable
modes. Additionally, we will repeat these calculations for all five models with the only 
difference that the axial magnetic field equals zero $(\alpha_2=0.0)$. Since we are only 
considering axisymmetric perturbations, these last calculations yield instabilities which 
can not have a magneto-rotational nature, and are purely convective instabilities. 
Because the axial magnetic fields are weak, comparing the growth rates
of a model with or without the axial magnetic field component provides 
a quantitative measure of the magneto-rotational contribution 
of an unstable mode in the interior of an accretion disk.

\begin{table}
\center
\vskip \baselineskip
\begin{tabular}{|c|c|c|c|} \hline 
& & & \\
Sim & $\epsilon$ & $\beta$ & $N^2/\Omega^2$ \\
& & & \\ \hline
& & & \\
A & 1.0 & 1000.0 & 600.0 \\
& & & \\
B & 0.94 & 1000.0 & 4.5\\
& & & \\
C & 0.88 & 1000.0 & 2.0 \\
& & & \\
D & 0.845 & 1000.0 & 1.5 \\
& & & \\
E & 1.0 & 20.0 & 12.0 \\
& & & \\
F & 1.0 & 10.0 & 6.0 \\
& & & \\
\hline 
\end{tabular}
\bigskip
\caption{Parameters of accretion disk models: all the models considered
have  $\alpha_1 =300$, $a=-3$ and $\alpha_2=1$ ($\alpha_2=0$) for models 
with (without) axial magnetic field.}
\end{table}

\begin{figure*}
\centering
\includegraphics[width=16cm]{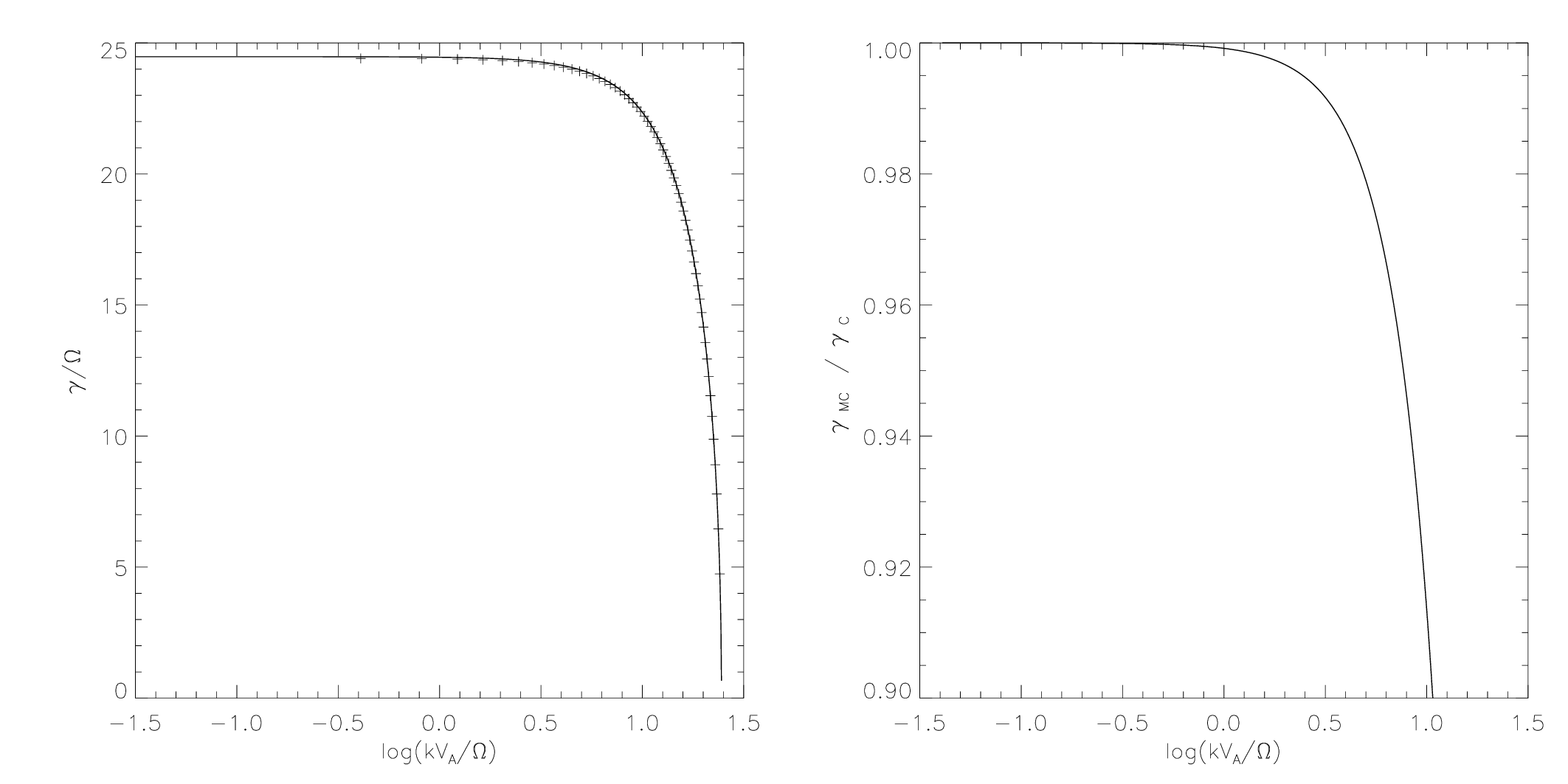}
\caption{Left panel: dimensionless growth rate $\gamma/\Omega$ as a 
function of dimensionless wavenumber $kv_{\rm A}/\Omega$ for MHD equilibrium
A (see table 1). The solid line corresponds with the solution 
from Narayan et al. (\cite{NY}). The crosses correspond to the solution
using the code LODES. Right panel: ratio of the growth rate $\gamma_{\rm MC}$ with 
($\alpha_2=1.0$) and the growth rate $\gamma_{\rm C}$ without ($\alpha_2=0.0$) 
an axial magnetic field component as a  function of the dimensionless wavenumber 
$kv_{\rm A}/\Omega$.
}
\label{Pap2Fig4}
\end{figure*}

\begin{figure*}
\centering
\includegraphics[width=16cm]{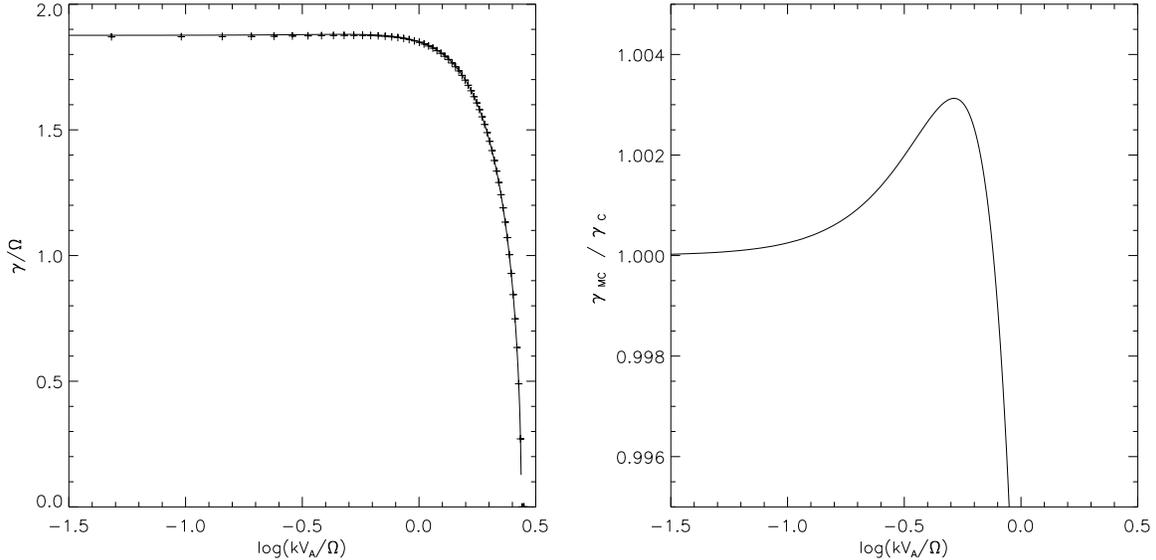}
\caption{Left panel: dimensionless growth rate $\gamma/\Omega$ as a 
function of dimensionless wavenumber $kv_{\rm A}/\Omega$ for MHD equilibrium
B (see table 1). The solid line corresponds with the solution 
from Narayan et al. (\cite{NY}). The crosses correspond to the solution
using the code LODES. Right panel: ratio of the growth rate $\gamma_{\rm MC}$ with 
($\alpha_2=1.0$) and the growth rate $\gamma_{\rm C}$ without ($\alpha_2=0.0$) 
an axial magnetic field component as a  function of the dimensionless wavenumber 
$kv_{\rm A}/\Omega$.
}
\label{Pap2Fig5}
\end{figure*}

\subsection{Weakly magnetized disks: from sub-Keplerian to near-Keplerian disks}

In this subsection, we will investigate the influence of the parameter $\epsilon$
on the nature of the instabilities in weakly magnetized ($\beta =1000$) MHD 
equilibria (models A-D, Table 1).  We will present the growth rates of the instabilities in
our model as a function of the axial wavenumber, using the same scaling
as Narayan et al. (\cite{NQIA}).

Figure~\ref{Pap2Fig4} shows our results for model A, which has a corresponding
value of the ratio $N^2/\Omega^2 = 600$ (model A, Table 1). The left panel shows a 
comparison between our calculations for the growth rate of the instabilities (crosses) 
and the calculation from Narayan et al. (\cite{NQIA}) (solid line). There is perfect agreement between both
results. The right panel shows the ratio of the growth rate of model
A with an axial magnetic field ($\alpha_2=1.0$) with respect to model A
without an axial magnetic field ($\alpha_2=0.0$). 
The model without axial magnetic field excludes the magneto-rotational 
mechanism completely.
One can see that the ratio never exceeds unity, implying that there is no
enhancement of the growth rate due to the magneto-rotational mechanism.
Furthermore, we observe in the left and right panel that there is no
observed maximum value of the growth rate at 
${\rm log}(kv_{\rm A}/\Omega)\; \approx \; 0.0 $, a feature also associated
with the magneto-rotational instability. Therefore, the observed 
instabilities have a (nearly) completely convective nature in this particular
disk model, as expected from the very high value of $N^2/\Omega^2$.

Figure~\ref{Pap2Fig5} shows our results for model B, which has a corresponding
value of the ratio $N^2/\Omega^2 = 4.5$ (model B, Table 1). One can see in 
the left panel that the growth rate from our calculations (crosses) again matches the 
solution for the growth rate (solid line) from Narayan et al.(\cite{NQIA}),
despite the fact that we consider an overall weak, but predominantly toroidal
magnetic field.
The right panel of figure~\ref{Pap2Fig5} shows the ratio of the growth rate of model B 
with axial magnetic field ($\alpha_2=1.0$), with respect to model B without an axial 
magnetic field component ($\alpha_2=0.0$). It is observed that there is an overall
small contribution (less than 1 percent) to the growth rates by the magneto-rotational 
mechanism. This indicates that the unstable modes have a dominant convective nature
in this model, like in model A.

\begin{figure*}
\centering
\includegraphics[width=16cm]{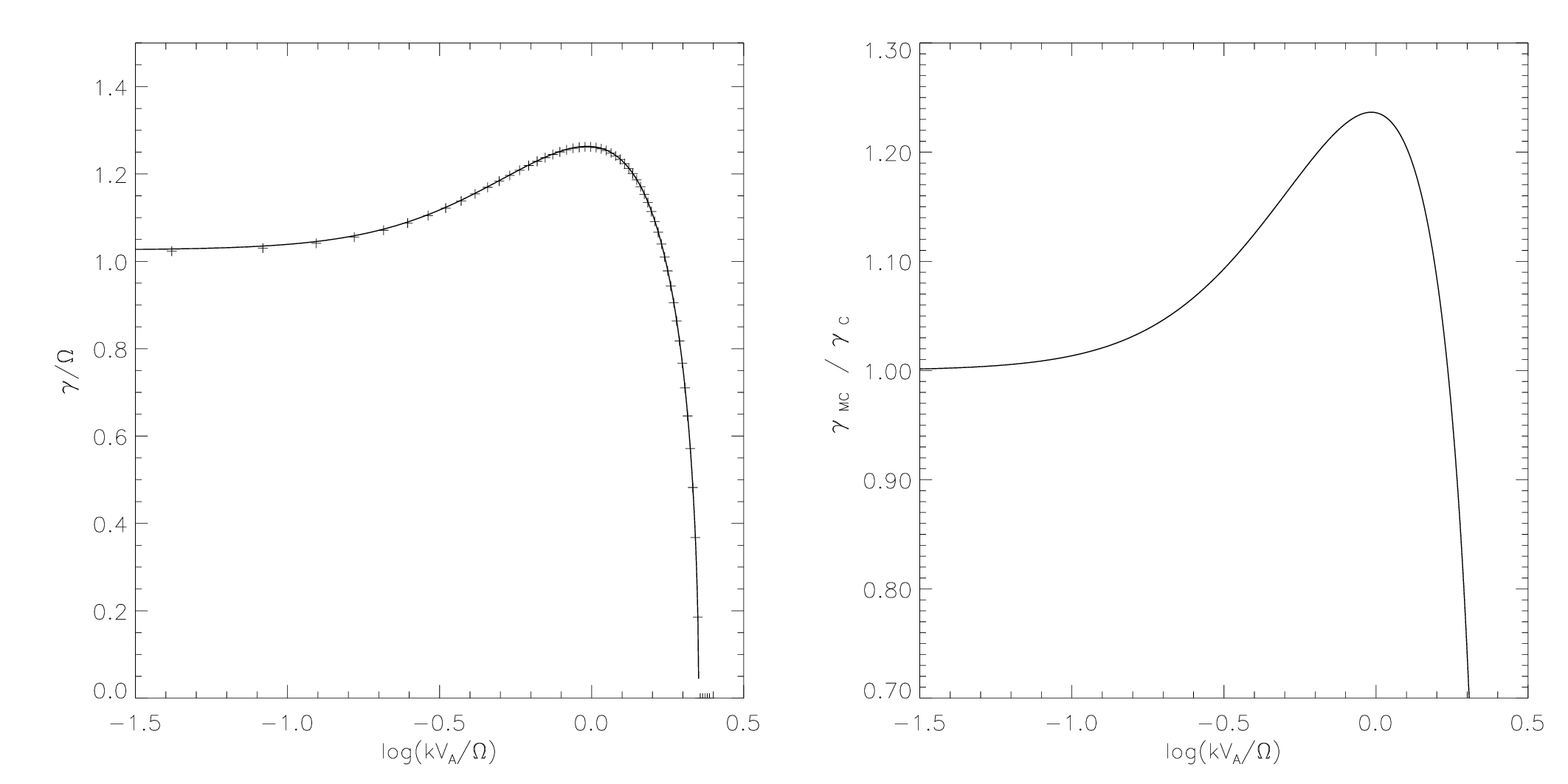}
\caption{Left panel: dimensionless growth rate $\gamma/\Omega$ as a 
function of dimensionless wavenumber $kv_{\rm A}/\Omega$ for MHD equilibrium
C (see table 1). The solid line corresponds with the solution 
from Narayan et al. (\cite{NY}). The crosses correspond to the solution
using the code LODES. Right panel: ratio of the growth rate $\gamma_{\rm MC}$ with 
($\alpha_2=1.0$) and the growth rate $\gamma_{\rm C}$ without ($\alpha_2=0.0$) 
an axial magnetic field component as a  function of the dimensionless wavenumber 
$kv_{\rm A}/\Omega$.
}
\label{Pap2Fig6}
\end{figure*}
\begin{figure*}
\centering
\includegraphics[width=16cm]{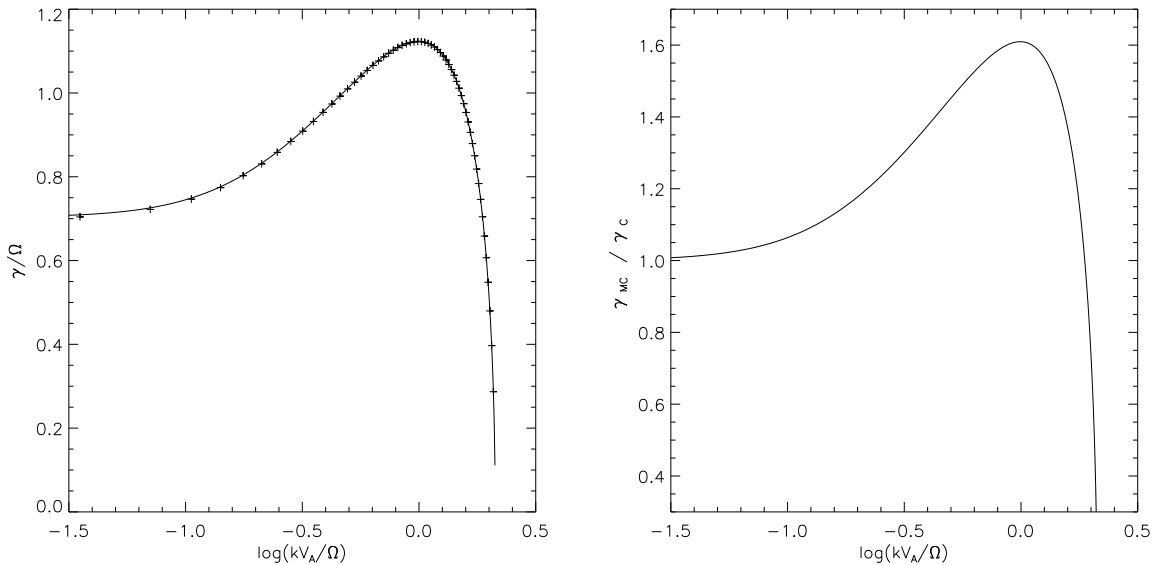}
\caption{Left panel: dimensionless growth rate $\gamma/\Omega$ as a 
function of dimensionless wavenumber $kv_{\rm A}/\Omega$ for MHD equilibrium
D (see table 1). The solid line corresponds with the solution 
from Narayan et al. (\cite{NY}). The crosses correspond to the solution
using the code LODES. Right panel: ratio of the growth rate $\gamma_{\rm MC}$ with 
($\alpha_2=1.0$) and the growth rate $\gamma_{\rm C}$ without ($\alpha_2=0.0$) 
an axial magnetic field component as a  function of the dimensionless wavenumber 
$kv_{\rm A}/\Omega$.
}
\label{Pap2Fig7}
\end{figure*}

In model C, the thickness of the disk model has been reduced further,
such that the obtained
ratio $N^2/\Omega^2\simeq 2.0$ equals the maximum value Narayan et al.
(\cite{NQIA}) have considered in their paper.
The left panel of Figure~\ref{Pap2Fig6} shows the calculated growth 
rates (crosses) of this MHD equilibrium, and there is a perfect 
match with the solution (solid line) from Narayan et al.(\cite{NQIA}). 
The right panel of Figure~\ref{Pap2Fig6} shows the ratio of the growth 
rate of model C with axial magnetic field ($\alpha_2=1.0$) with respect 
to model C without an axial magnetic field component ($\alpha_2=0.0$). 
The contribution from the magneto-rotational mechanism is significant, 
of order 25 percent around values of the wavenumber where the growth 
rate has its maximum: this increase indicates that the 
unstable modes in this range of wavenumbers are significantly amplified 
by the magneto-rotational mechanism.
Furthermore, model C shows that for axial wavenumbers 
${\rm log}(kv_{\rm A}/\Omega) < -1.0$, the contribution of the 
magneto-rotational mechanism has decreased to a few percent. This 
indicates that the unstable modes in this range of wavenumbers have a 
dominant convective nature. This conclusion on the long wavelength regime
was also made by Narayan et al. (\cite{NQIA}) in their analysis.

\begin{figure*}
\centering
\includegraphics[width=16cm]{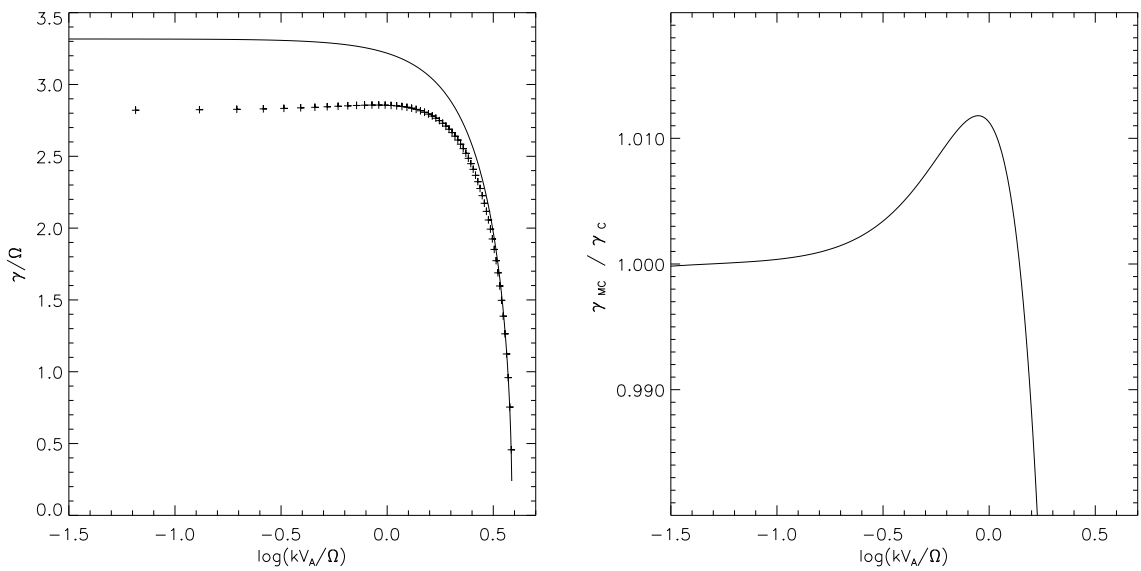}
\caption{Left panel: dimensionless growth rate $\gamma/\Omega$ as a 
function of dimensionless wavenumber $kv_{\rm A}/\Omega$ for MHD equilibrium
E (see table 1). The solid line corresponds with the solution 
from Narayan et al. (\cite{NY}). The crosses correspond to the solution
using the code LODES. Right panel: ratio of the growth rate $\gamma_{\rm MC}$ with 
($\alpha_2=1.0$) and the growth rate $\gamma_{\rm C}$ without ($\alpha_2=0.0$) 
an axial magnetic field component as a  function of the dimensionless wavenumber 
$kv_{\rm A}/\Omega$.
}
\label{Pap2Fig8}
\end{figure*}
\begin{figure*}
\centering
\includegraphics[width=16cm]{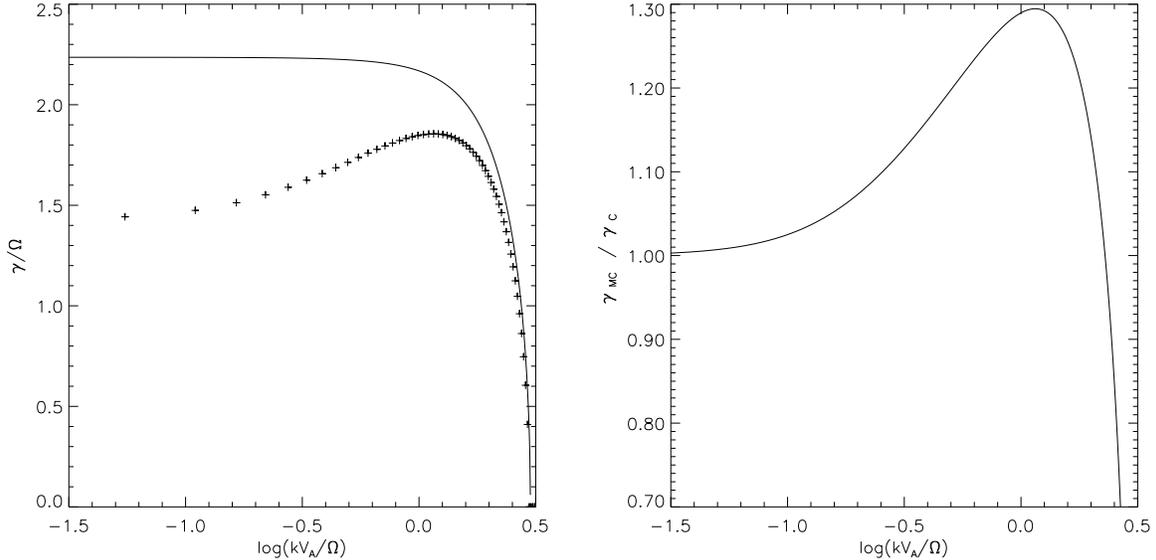}
\caption{Left panel: dimensionless growth rate $\gamma/\Omega$ as a 
function of dimensionless wavenumber $kv_{\rm A}/\Omega$ for MHD equilibrium
F (see table 1). The solid line corresponds with the solution 
from Narayan et al. (\cite{NY}). The crosses correspond to the solution
using the code LODES. Right panel: ratio of the growth rate $\gamma_{\rm MC}$ with 
($\alpha_2=1.0$) and the growth rate $\gamma_{\rm C}$ without ($\alpha_2=0.0$) 
an axial magnetic field component as a  function of the dimensionless wavenumber 
$kv_{\rm A}/\Omega$.
}
\label{Pap2Fig9}
\end{figure*}

Finally, Figure~\ref{Pap2Fig7} shows our results for model D, which has 
a corresponding value of the ratio $N^2/\Omega^2 = 1.5$ (see Table 1). 
The left panel of Figure~\ref{Pap2Fig7} shows the calculated growth 
rates (crosses) of this MHD equilibrium, in agreement with the solution 
(solid line) from Narayan et al.(\cite{NQIA}). 
The right panel of Figure~\ref{Pap2Fig7} shows the ratio of the growth 
rate of model D with axial magnetic field ($\alpha_2=1.0$) with respect 
to model D without an axial magnetic field component ($\alpha_2=0.0$). 
The contribution from the magneto-rotational mechanism is increasing up
to 60 percent around values of the wavenumber where the growth rate has 
its maximum. For axial wavenumbers ${\rm log}(kv_{\rm A}/\Omega) < -1.0$, 
the contribution of the magneto-rotational mechanism has a maximum of
order 5 percent, which indicates their dominant convective nature.  

The models C and D discussed above have the same ratio of $N^2/\Omega^2$
as the models discussed by Narayan et al.(\cite{NQIA}). Our conclusions
confirm those by Narayan et al.(\cite{NQIA}) for the unstable long-wavelength
modes ($kV_{\rm A}\ll \Omega$) in these two models: they are clearly convective modes.  
However, the wavelength modes in these models with $kV_{\rm A}\sim \Omega$ are shown
to be maximally amplified by the magneto-rotational mechanism: the amplification
is increasing as the ratio $N^2/\Omega^2$ is decreasing. However, these modes seem to 
be determined by both the convective and the magneto-rotational mechanism. Finally, 
for modes with $kV_{\rm A} > \Omega$, the amplification of the growth rate is shown to 
be decreasing, due to the restoring magnetic tension force.

\subsection{From weakly magnetized disks to disks close to equipartition}

In this subsection we investigate the influence of the magnetization on
the nature of the instabilities in our considered MHD equilibria (models
E \& F, Table 1).

Figure~\ref{Pap2Fig8} shows our results for model E. The considered MHD 
equilibrium has the same parameters as model A from the previous subsection, except 
this one is closer to equipartition ($\beta = 20$). Due to the stronger magnetic field, 
this equilibrium has a lower value of the ratio $N^2/\Omega^2$ (see Table 1). 
Our calculations (crosses) and the calculation from 
Narayan et al. (\cite{NQIA}) (solid line) deviate from each other, since
the equation used by Narayan et al. (\cite{NQIA}) is
not valid for this model. This is because the toroidal magnetic field strength
can not be neglected anymore in the linear analysis, a component which was
not taken into account in the equation used by Narayan et al. (\cite{BSKG}).
Contrary to the model A, this model shows a maximum value of the growth
rate, close to the typical value of the  magneto-rotational instability,
i.e. ${\rm log}(kv_{\rm A}/\Omega)\; \approx \; 0.0 $.
The right panel shows the ratio of the growth rate of model E with axial 
magnetic field ($\alpha_2=1.0$) with respect to model E
without an axial magnetic field component ($\alpha_2=0.0$). There is a
slight enhancement observed close to the maximum value of the growth rate:
this indicates that the magneto-rotational mechanism starts to contribute 
to the growth rate of unstable modes in this region. Notice that the 
enhancement is still very weak, of order 1 percent (right panel Figure~\ref{Pap2Fig8}).

Finally, figure~\ref{Pap2Fig9} shows our results for model F. In this MHD equilibrium
the magnetization is close to equipartition ($\beta =10.0$), with a corresponding value 
$N^2/\Omega^2=6.0$ (see Table 1). We notice that we do not consider a disk in exact
equipartition (i.e. $\beta =1.0$), because this model would not allow for convective instabilities
(i.e. $N^2 / \Omega^2 < 1$ as can be seen in Figure 3). The left panel of 
Figure ~\ref{Pap2Fig9} shows that the approximation from Narayan et al. (\cite{NQIA}) (solid line) 
is no longer valid. Taking into account the toroidal magnetic field component, our calculations show 
a distinct maximum which occurs at the typical value ${\rm log}(kv_{\rm A}/\Omega) =0.0$ 
for magneto-rotational instabilities. 
The right panel of figure~\ref{Pap2Fig9} shows the ratio of the growth rate of model F 
with axial magnetic field ($\alpha_2=1.0$) with respect to model F without an axial 
magnetic field component ($\alpha_2=0.0$). It can be observed that in the region around 
the maximum of the growth rate, the contribution of the magneto-rotational mechanism is
of order 30 percent. This increase indicates that the unstable modes in this range of
wavenumbers are significantly amplified by the magneto-rotational mechanism. However, for 
axial wavenumbers ${\rm log}(kv_{\rm A}/\Omega) < -1.0$, the contribution of the 
magneto-rotational mechanism has decreased to a few percent. This indicates that the unstable 
modes in this range of wavenumbers have a dominant convective nature.

\section{Conclusions}

We calculated growth rates of instabilities for MHD equilibria of
magnetized accretion disks. The disks models were taken in the cylindrical 
limit and included both toroidal and axial magnetic field components. 
Due to the presence of a toroidal magnetic field component, the
unstable modes we find are strictly speaking {\it overstable} modes 
(see e.g. Blokland et al., \cite{BSKG}).
The parameters of the model in our calculations were taken such that 
the axial wavenumber of the most unstable modes were larger than the
inversed scale height $H$. Therefore the analysis can be used as an
approximation for a stability analysis of the interior of an accretion
disk. Our models considered both weakly magnetized disks, as 
well as disks which are close to equipartition. All calculations
were performed with a restriction to axisymmetric perturbations ($m=0$).

Our most important results are summarised below:
\begin{itemize}
\item
We have considered models of thick accretion disks, which are 
subject to convective magneto-rotational instabilities,
and we quantified the contribution of the MRI mechanism of all
the instabilities considered. 
\item
Our calculations are diverging from the solutions from Narayan
et al.(\cite{NQIA}), when the disk models are becoming significantly
magnetized. The deviations are contributed to the toroidal magnetic field 
component, which was neglected in the derivation of the equation for the
growth rate used by Narayan et al.(\cite{NQIA}). 
\item
Our calculations show that the contribution from the magneto-rotational
mechanism to the growth rate becomes significant as the disks get close 
to equipartition. The contribution peaks for modes near the maximum value
of the growth rate.        
\item
Our calculations for the growth rates agree with the calculations from 
Narayan et al. (\cite{NQIA}) for weakly magnetized accretion disk models,
even though we include dominant toroidal field components.  
\item
Our calculations show that the contribution from the magneto-rotational
mechanism to the growth rate becomes significant as the scale-height $H$
is decreased. This contribution peaks for modes near the maximum value
of the growth rate.
\item
All our calculations show that the contribution from the magneto-rotational
mechanism to the growth rate becomes neglible for unstable modes with
${\rm log}(kv_{\rm A}/\Omega) < -1.0$, which are safely interpreted as
purely convective modes in all cases.
\end{itemize}
\begin{acknowledgements}
JWB and RK carried out this work within the framework of the
European Fusion Programme, and it is supported by the European
Communities under the contract of Association between EURATOM/FOM. Views and
opinions expressed herein do not necessarily reflect those of the European 
Commission. EvdS did this research in the FOM projectruimte on 
`Magnetoseismology of
accretion disks', a collaborative project between R. Keppens (FOM Institute
Rijnhuizen, Nieuwegein) and N. Langer (Astronomical Institute Utrecht). 
This work is part of the research programme of the `Stichting voor 
Fundamenteel Onderzoek der Materie (FOM)', which is financially supported by 
the `Nederlandse Organisatie voor Wetenschappelijk Onderzoek (NWO)'.
\end{acknowledgements}

\end{document}